# An electrically-driven Carbon nanotube-based plasmonic laser on Silicon

Ke Liu[1, 2], Behrouz Movahhed Nouri[3], Elham Heidari[3], Hamed Dalir[3] and Volker J. Sorger[1,3*]

[1]Department of Electrical and Computer Engineering, George Washington University, Washington, DC 20052, USA
[2]The Key Laboratory of Optoelectronics Technology, Ministry of Education, Beijing University of Technology, Beijing 100124, China
[3]Optelligence LLC, 10703 Marlboro Pike, Upper Marlboro, MD, 20772, USA
[*]sorger@gwu.edu

**Abstract:** Photonic signal processing requires efficient on-chip light sources with higher modulation bandwidths. Today's conventional fastest semiconductor diode lasers exhibit modulation speeds only on the order of a few tens of GHz due to gain compression effects and parasitic electrical capacitances. Here we theoretically show an electrically-driven Carbon nanotube (CNT)-based laser utilizing strong light-matter-interaction via monolithic integration into Silicon photonic crystal nanobeam (PCNB) cavities. The laser is formed by single-walled CNTs inside a combo-cavity consisting of both a plasmonic metal-oxide-semiconductor hybrid mode embedded in the one dimensional PCNB cavity. The emission originates from interband recombinations of electrostatically-doped nanotubes depending on the tubes' chirality towards matching the C-band. Our simulation results show that the laser operates at telecom frequencies resulting in a power output > 3 (100) μW and > 100 (1000)'s GHz modulation speed at 1× (10×) threshold. Such monolithic integration schemes provide an alternative promising approach for light source in future photonic integrated circuits.

**References and links**

1. P. Avouris, M. Freitag, and Vasili Perebeinos, "Carbon-nanotube photonics and optoelectronics," Nat. Photonics **2**, 341-350 (2008).
2. M. Bansal, R. Srivastava, C. Lal, M.N. Kamalasanan, and L.S. Tanwar, "Carbon nanotube-based organic light emitting diodes," Nanoscale **1**, 317-330 (2009).
3. X. Wang, L. Zhang, Y. Lu, H. Dai, Y. K. Kato, and Eric Pop, "Electrically driven light emission from hot single-walled carbon nanotubes at various temperatures and ambient pressures," Appl. Phys. Lett. **91**, 261102 (2007).
4. E. Gaufrès, N. Izard, X. Le Roux, D. Marris-Morini, S. Kazaoui, E. Cassan, and L. Vivien, "Optical gain in carbon nanotubes," Appl. Phys. Lett. **96**, 231105 (2010).
5. Y. Miyauchi, M. Iwamura, S. Mouri, T. Kawazoe, M. Ohtsu, andK. Matsuda, "Brightening of excitons in carbon nanotubes on dimensionality modification," Nat. Photonics **7**, 715-719 (2013).
6. T. Mueller, M. Kinoshita, M. Steiner, V. Perebeinos, A.A. Bol, D.B. Farmer, and P. Avouris, "Efficient narrow-band light emission from a single carbon nanotube p-n diode," Nat. Nanotechnol. **5**, 27-31 (2010).
7. S. Wang, Q. Zeng, L. Yang, Z. Zhang, Z. Wang, T. Pei, L. Ding, X. Liang, M. Gao, Y. Li, and L.M. Peng, "High-performance Carbon nanotube light-emitting diodes with asymmetric contacts," Nano Lett. **11** (1), 23-29 (2011).
8. E. Gaufrès, N. Izard, A. Noury, X. Le Roux, G. Rasigade, A. Beck, and L.Vivien, "Light emission in Silicon from Carbon nanotubes," ACS Nano **6** (5), 3813-3819 (2012).
9. S. Khasminskaya, F. Pyatkov, B.S. Flavel, W.H. Pernice, and R.Krupke, "Waveguide-integrated light-emitting Carbon nanotubes," Adv. Mater. **26**, 3465-3472 (2014).
10. S. Bahena-Garrido, N. Shimoi, D. Abe, T. Hojo, Y. Tanaka, and K. Tohji, "Plannar light source using a phosphor screen with single-walled carbon nanotubes as field emitters," Rev. Sci. Instrum. **85**, 104704 (2014).

11. D. Yu, H. Liu, L.M. Peng, and S. Wang, "Flexible light-emitting devices based on chirality-sorted semiconducting carbon nanotube films," ACS Appl. Mater. Interfaces **7** (6), 3462-3467 (2015).
12. G.H. Duan, C. Jany, A.L. Liepvre, A. Accard, M. Lamponi, D. Make, P. Kaspar, G. Levaufre, N. Girard, F. Lelarge, J.M. Fedeli, A. Descos, B.B. Bakir, S. Messaoudene, D. Bordel, S. Menezo, G.D. Valicourt, S. Keyvaninia, G. Roelkens, D.V. Thourhout, D.J. Thomson, F.Y. Gardes, and G.T. Reed, "Hybrid III-V on Silicon lasers for photonic integrated circuits on Silicon," IEEE J. Sel. Top. Quantum. Electron. **20** (4), 6100213 (2014).
13. S. Wu, S. Buckley, J.R. Schaibley, L. Feng, J. Yan, D.G. Mandrus, F. Hatami, W. Yao, J. Vučković, A. Majumdar, and X. Xu, "Monolayer semiconductor nanocavity lasers with ultralow thresholds," Nature **520**, 69-72(2015).
14. K. Ding, M. T. Hill, Z.C. Liu, L.J. Yin, P.J. van Veldhoven, C.Z. Ning, "Record performance of electrical injection sub-wavelength metallic-cavity semiconductor lasers at room temperature," Opt. Express **21**, 4728-4733 (2013).
15. R.F. Oulton, V.J. Sorger, T. Zentgraf, R.M. Ma, C. Gladden, L. Dai, G. Bartal, and X. Zhang, "Plasmon lasers at deep subwavelength scale," Nature **461**, 629-632 (2009).
16. V.J. Sorger, N. Pholchai, E. Cubukcu, R.F. Oulton, P. Kolchin, C. Borschel, M. Gnauck, C. Ronning, and X. Zhang, "Strongly enhanced molecular fluorescence inside a nanoscale waveguide gap," Nano Lett. **11** (11), 4907-4911 (2011).
17. K. Liu, C.R. Ye, S. Khan, and V.J. Sorger, "Review and perspective on ultra-fast and wavelength-size electro-optic modulators," Laser Photon. Rev. **9**(2), 172-194 (2015).
18. K.Y. Jeong, Y.S. No, Y. Hwang, K. S. Kim, M.K. Seo, H.G. Park, and Y.H. Lee, "Electrically driven nanobeam laser," Nat. Commun. **4**, 2822 (2013).
19. K. Liu and V.J. Sorger, "Enhanced interaction strength for a square plasmon resonator embedded in a photonic crystal cavity", J. Nanophotonics **9**(1), 093790 (2015).
20. A.R.M. Zain, N.P. Johnson, M. Sorel, and R.M. De La Rue, "Ultra high quality factor one dimensional photonic crystal/photonic wire micro-cavities in silicon-on-insulator (SOI)," Opt. Express **16** (16), 12084-12089 (2008).
21. Q. Quan and M. Loncar, "Deterministic design of wavelength scale, ultra-high Q photonic crystal nanobeam cavities," Opt. Express **19** (19), 18530-18542 (2011).
22. J.T. Robinson, C. Manolatou, L.Chen, and M. Lipson, "Ultrasmall mode volumes in dielectric optical microcavities," Phys. Rev. Lett. **95**, 143901 (2005).
23. E. M. Purcell, "Spontaneous emission probabilities at radio frequencies," Phys. Rev. **69**(1-2), 681 (1946).
24. L.W. Luo, G.S. Wiederhecker, J. Cardenas, C. Poitras, and M. Lipson, "High quality factor etchless silicon photonic ring resonators," Opt. Express **19**(7), 6284-6289 (2011).
25. T. Yoshie, J. Vučković, A. Scherer, H. Chen, and D. Deppe, "High quality two-dimensional photonic crystal slab cavities," Appl. Phys. Lett. **79**, 4289 (2001).
26. E.J.R. Vesseur, F.J. García de Abajo, and A. Polman, "Broadband Purcell enhancement in plasmonic ring cavities," Phys. Rev. B **82**, 165419 (2010).
27. R.M. Ma, R.F. Oulton, V.J. Sorger, and X. Zhang, "Plasmon lasers: coherent light source at molecular scales," Laser Photon. Rev. **7** (1), 1-21 (2013).
28. C.Y. Lu, C.Y. Ni, M. Zhang, S. L. Chuang, and D.H. Bimberg, "Metal-cavity surface-emitting microlasers with size reduction: theory and experiment," IEEE J. Sel. Top. Quantum. Electron. **19** (5), 1701809 (2013).
29. J.M. Marulanda and A.Srivastava, "Carrier density and effective mass calculations in carbon nanotubes," Phys. Stat. Sol. (b) **245** (11), 2558-2562 (2008).
30. P. Lalanne, C. Sauvan, and J.P. Hugonin, "Photon confinement in photonic crystal nanocavities," Laser Photon. Rev. **2**(6), 514-526 (2008).
31. D.A. Genov, R.F. Oulton, G. Bartal, and X. Zhang, "Anomalous spectral scaling of light emission rates in low-dimensional metallic nanostructures," Phys. Rev. B **83**, 245312 (2011).
32. G.S. Tulevski, A.D. Franklin, D. Frank, J.M. Lobez, Q. Cao, H.Park, A. Afzali, S.J. Han, J.B. Hannon, and W. Haensch, "Toward high-performance digital logic technology with Carbon nanotubes," ACS Nano 8(9), 8730-8745 (2014).
33. M.S. Arnold, A.A. Green, J.F. Hulvat, S.I. Stupp, and M.C. Hersam, "Sorting carbon nanotubes by electronic structure using density differentiation," Nat. Nanotechnol. 1, 60-65 (2006).
34. J.A. Fagan, M.L. Becker, J.H. Chun, P.T. Nie, B.J. Bauer, J.R. Simpson, A. Hight-Walker, and E.K. Hobbie, "Centrifugal length separation of Carbon nanotubes," Langmuir 24, 13880-13889 (2008).
35. G.S. Tulevski, A.D. Franklin, and A. Afzali, "High purity isolation and quantification of semiconducting Carbon nanotubes via column chromatography," ACS Nano 7(4), 2971-2976 (2013).
36. C.Y. Khripin, J.A. Fagan, and M. Zheng, "Spontaneous partition of Carbon nanotubes in polymer-modified aqueous phases," J. Am. Chem. Soc. 135 (18), 6822-6825 (2013).
37. S. Shekhar, P. Stokes, and S. I. Khondaker, "Ultrahigh density alignment of Carbon nanotube arrays by dielectrophoresis," ACS Nano 5(3), 1739-1746 (2011).

38. A. Vijayaraghavan, S. Blatt, D. Weissenberger, M. Oron-Carl, F. Hennrich, D. Gerthsen, H. Hahn, and R. Krupke, "Ultra-large-scale directed assembly of single-walled carbon nanotube devices," Nano Lett. 7(6),1556-1560 (2007).
39. A. Vijayaraghavan, F. Hennrich, N. Stürzl, M. Engel, M. Ganzhorn, M. Oron-Carl, C.W. Marquardt, S. Dehm, S. Lebedkin, M.M. Kappes, and R. Krupke, "Toward single-chirality carbon nanotube device arrays," ACS Nano 4 (5), 2748-2754 (2010).
40. Y. Che, H. Chen, H. Gui, J. Liu, B. Liu and C. Zhou, "Review of carbon nanotube nanoelectronics and macroelectronics," Semicond. Sci. Technol. 29, 073001 (2014).
41. S.J. Tans, A.R.M. Verschueren, and C. Dekker, "Room-temperature transistor based on a single carbon nanotube," Nature 393 (6680), 49-52 (1998).
42. R. Martel, T. Schmidt, H.R. Shea, T. Hertel, and P. Avouris, "Single- and multi-wall carbon nanotube field-effect transistors," Appl. Phys. Lett. 73 (17), 2447-2449 (1998).
43. J.A. Misewich, R. Martel, Ph. Avouris, J.C. Tsang, S. Heinze, and J. Tersoff, "Electrically induced optical emission from a carbon nanotube FET," Science 300(5620), 783-786 (2003).
44. P. Rai, N. Hartmann, J. Berthelot, J. Arocas, G. Colas des Francs, A. Hartschuh, and A. Bouhelier, "Electrical excitation of surface plasmons by an Individual Carbon nanotube transistor," Phys. Rev. Lett. 111, 026804 (2013).

## 1. Introduction

Semiconducting single-walled Carbon nanotubes (CNTs) are being recently explored for photonic integrated circuits due to their unique electronic and optical properties [1, 2]. Light amplification in Carbon nanotubes was experimentally demonstrated in the near-infrared wavelength range at cryo [3] and room temperatures [4], as a single photon emitter through dimensionality modification [5], by tuning the direct band-gap, controlling excitonic recombinations, and enabling exciton radiatively-decaying. Device examples of light emission from CNTs have previously demonstrated a p-n diode [6, 7], tube to waveguide-coupling [8, 9], flat plane-emission panels [10], and flexible light-emitting sources [11]. However, CNTs-based laser devices operating at a telecom wavelength, which are desired for on-chip optical interconnects, are not reported to date.

Carbon nanotubes sorting (i.e. semiconducting, diameter, or single chirality) and placement (i.e. position precisely at a predefined location and orientation) are two of the key challengers in the development of CNT-based optoelectronic devices [Tule14]. For the sorting, surfactant-based separation solutions are utilized counting on CNTs post-growth processing through electronic type and diameter, such as density gradient ultracentrifugation technique [33, 34], showing a semiconducting purity of >99%, column chromatography method [35] due to metallic and semiconducting CNTs' moving at different rates for separation. Other types of polymer extractions techniques are also effective in sorting CNTs, for instance, large (1.2~1.5 nm) and small (0.6 ~ 1.0 nm) diameter CNTs from solution can be successfully extracted by the addition of water-soluble polymers [36]. In terms of the CNTs replacement, up to date two different placement strategies are classified depending on CNTs' growth, purification, and placement accomplished either in one step or in three completely separated process [32]. The aim is to enable the sorted CNTs to transfer on a complementary metal-oxide-semiconductor (MOS) compatible substrate. Among these methods, directed assembly using dielectrophoresis including alternating current [37], and radio frequency [38] exhibits a promising method for alignment of CNTs between metal contacts, where a large scale assembly of individual CNTs can bridge each electrode pair.

Compared to conventional bulk Silicon MOS field-effect transistors (FET), CNTFET exhibits superior performance due to its transconductance and drive currents by a factor of a few per unit width, making it an attractive alternative to Silicon [40]. CNT field-effect transistors were first demonstrated as early as 1998 [41, 42]. With applying proper bias scheme these CNTFET create p-n junction and behave as diode device, operating more closely as rectifiers with a forward bias and limited current flow with the reverse direction. However, here we focus on electrically-induced light emission (i.e. electroluminescence) with a gain option from carbon nanotubes for laser applications. Different aspects of light emission mechanisms depend on CNT device structures, such as using various gate configurations (e.g. bottom gate, and top split-gate). Optical emission, which originates from radiative recombination of electrons and holes simultaneously injected into the undoped nanotube, was first observed from a three-terminal ambipolar type CNTFET having with a forward-biased p-n junction [43]. However, two-terminal CNT-based light emitting diodes are usually the basic building block in modern optoelectronic circuits due to their significant advantages (e.g. lower power consumption and cost, relative simpler drive circuitry) as light sources [7], which is used in this work. Basically light emission from CNT devices involves radiative combination of electron and holes, either as free carriers or bound in the form of excitons.

A laser is constructed from three principal parts including gain medium, optical cavity, and pump source (either optical or electrical). The observation of optical gain in semiconducting single-walled CNTs is of great importance to the proper design of laser devices. Fortunately, the significant optical gain in (8, 7) single-walled CNTs embedded in host polymer thin film was experimentally demonstrated at a wavelength of 1.3 µm at room temperature [4], showing that carbon nanotubes are able to amplify light. Therefore, here a laser can be potentially obtained by inserting single-walled CNTs material (i.e. gain medium) into the optical cavity (e.g. photonic crystal nanobeam cavity for our case). Lasing effect may be achieved as the optical gain exceeds a threshold value determined by the cavity loss mechanism resulting from stimulated absorption and intrinsic loss.

With the aim to design a CNT-based laser, a significant challenge is the inherently small overlap factor between the tube's gain material with the optical mode, requiring light-matter interaction (LMI) enhancement techniques. Next we briefly outline some LMI options to be considered including one-dimensional (1-D) interference grating (i.e. distributed Bragg reflectors), photonic crystal, metal-clad, and plasmonic [12-15]. Regarding the latter, the metal-oxide-semiconductor (MOS) configuration can support a hybrid plasmon-polariton (HPP) waveguide mode, where the peak of the electric field intensity is mainly concentrated in the thin oxide gap, which can be collocated with the CNT gain material (i.e. placing the CNT inside the oxide gap) [16]. This mode provides synergies relating to photonic integration and active optoelectronics [17], such as enhanced LMIs via deep sub-diffraction limited modes, seamless access to semiconductors and integration with the Silicon-on-insulator (SOI)-platform for low loss routing. A 1-D photonic crystal nanobeam (PCNB) cavity operates as a Fabry-Perot-like resonator, offering optical confinement between Bragg mirrors consisting of a periodic array of air holes along the waveguide direction. For instance, an electrically driven, room-temperature 1-D PCNB laser with 0.35 $(\lambda/n)^3$ mode volume was demonstrated at a lasing wavelength of 1578 nm [18].In this work we aim to deploy strongly enhanced LMIs by using both the 1-D PCNB and the plasmonic MOS mode simultaneously towards realizing a high gain material-mode overlap for a CNT-based integrated nanolaser. We recently show a 44 times enhanced interaction strength for a square plasmon resonator with III-V materials embedded in a Silicon-based PCNB cavity, due to the highly compressed mode volume compared to the inline plasmon resonator without the cavity [19].

Towards enhancing the LMI between the CNTs and a cavity, we combine the MOS structure with the PCNB cavity and placing single-walled CNTs inside this combo-cavity. We theoretically show this approach for CNTs-based lasers to be seamlessly integrated into on-chip Silicon waveguides delivering potential high modulation bandwidth for planar chip architectures. Investigations of the cavity quality ($Q$) factor and Purcell factor, result in laser performance as derived by the light-matter interaction modified rate equations that outperforms classical laser devices.

## 2. Laser and cavity design

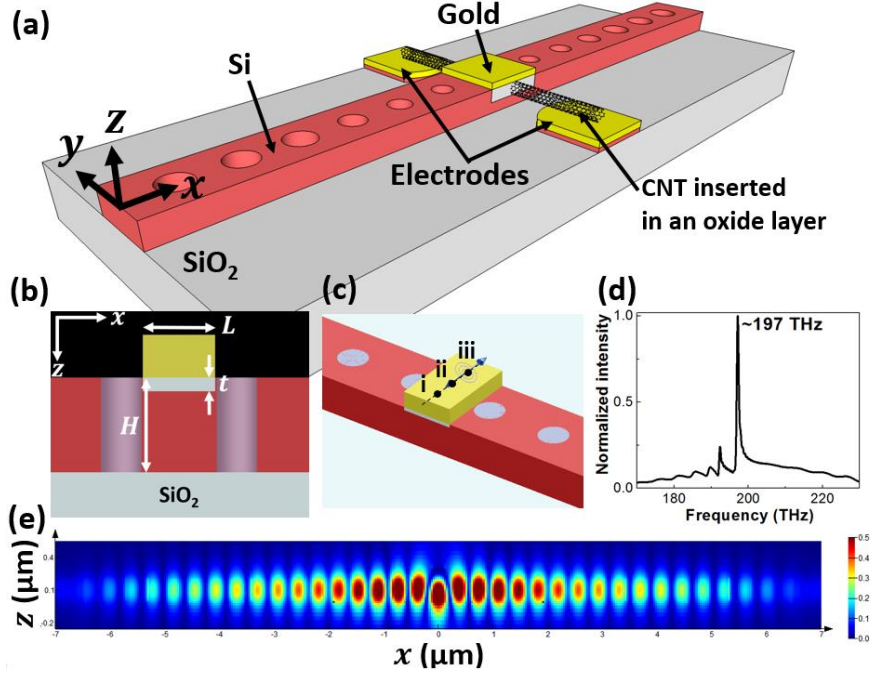

Fig. 1. (a) 3-D schematic structure of Carbon nanotube-based PCNB laser device with electrically driven scheme. A Gold-Oxide-Silicon stack forms a MOS configuration embedded in the center of the photonic crystal cavity, and ten single-walled CNTs parallel-aligned inside the oxide layer serves as active gain medium for light emission. Some physical parameters for the previous 1-D PCNB cavity design are unchanged such as hole period of $a$=380 nm, minimum hole spacing in the taper section of $a_{min}$ =350 nm, hole radius of $r$=0.2$a$, number of taper and mirror pairs of $n$=8, $m$=10, and waveguide width of 400 nm, which gives a cavity resonant wavelength of ~1520 nm. (b) Cross-sectional view of the plasmonic cavity section embedded in the PCNB cavity in the $xz$ plane. The Gold thickness is kept constant to 100 nm, and the total cavity height, $H$, is unchanged with 220 nm for the compatibility of commercially available SOI wafers. The oxide layer thickness, $t$, is varied from 0 to 50 nm for the design optimization. (c) Angled view of setting point-dipole excitation source with various positions and orientations for the modeling of single-walled CNTs emission. Three solid dots labeled by i, ii, iii indicate the positions of the dipole source placed along the $y$ direction within the oxide layer of the cavity center, and at each position the dipole source with three orientations (i.e. $x, y,$ and $z$) is used for the excitation, respectively, for the optimization of $Q$ and Purcell factor. (d) The resonant spectrum of the optimized MOS configuration embedded in the PCNB cavity, exhibiting a lasing frequency of ~197 THz (i.e. ~1522 nm). (e) Cross-section through a 3-D FDTD simulation of the electric-field distribution profile with intensity normalized between 0 and 0.5, showing the coupling of light from the CNT emitter into the PCNB cavity, as well as the propagation of light within the waveguide.

A high $Q$ 1-D PCNB cavity without the MOS structure is first designed at a target resonant wavelength of ~1550 nm. The design process of a 1-D PCNB cavity usually consists of engineering three elements [20, 21]: (1) the photonic crystal mirror, (2) the taper section, and (3) the cavity length. Here the cavity length of $L$=260 nm is optimized in our previous work [19], and the cavity height of $H$=220 nm is held constant for the compatibility of commercially available SOI wafers. A photonic ridge waveguide on SOI with the cross-section of height ($H$) 220 nm and width ($W$) 400 nm supporting a transverse-electric (TE) mode is deployed as a core building block for the laser design. The photonic crystal mirror and the taper parameters, including the hole period ($a$), hole radius ($r$), minimum hole spacing in the taper section ($a_{min}$), and number of taper and mirror pairs ($n, m$), are optimized for a highest $Q$ factor by sweeping $a, r$, and $a_{min}$. This design is performed by using a commercial software package FDTD Solutions distributed by Lumerical. The input of complex refractive indices (i.e. $n$ and $\kappa$) of Gold, SiO$_2$, and Si are taken from the solver's built-in material database. A reasonable high $Q$ cavity of $\sim 2 \times 10^4$ is found as $a$=380 nm, $a_{min}$=350 nm, $r$=0.2$a$, $n$=8, and $m$=10 for a cavity resonant wavelength of ~1550 nm.

Next we embed the plasmonic MOS mode into the PCNB cavity towards enhancing the LMI. Inside this high electric field of the combo-cavity we inserted 10 single-walled CNTs (the chiral number of (9,2), the diameter of ~1.0 nm, and the bandgap of ~0.85 eV) at the collocated with a thin oxide layer (Fig. 1a, b). Excitation of semiconducting CNTs can be done either optically or electrically. However, electrically pumping is more preferred for our structure since the metal pad forming the plasmonic mode can be conveniently used as a gate electrode to electrostatically dope the CNTs. Here the excitation of CNTs is considered driven electrically via a p-n junction at the nanotubes. Light created by spontaneous emission through electrons and holes recombination has a fixed polarization state along with long-axis of carbon nanotube [43, 44]. Hence we treat the generated light classically using electromagnetic point dipole source. The CNT emitters were first modeled by a dipole source with all 3 spatial positions and 9 polarization orientations (Fig. 1c). Among them, we found the $y$-polarized dipole source excitation is preferred for a PCNB cavity that is typically compatible with TE polarized light supported in the photonic SOI ridge waveguide. In addition, here single-walled CNTs are parallel-aligned along the $y$ coordinate axis inside the oxide layer, which physically meet the requirement of polarized electroluminescence emission in single-walled CNTs along the axis of the nanotube. The resulting electroluminescence of the nanotubes is generated in the thin oxide layer forming a hybrid HPP waveguide mode, which contributes to a PCNB cavity lasing mode. The transmission (reflection) spectrum was recorded at the output (input) port, respectively. At the resonant frequency of ~197 THz, showing ~60% transmitted light, we thus conclude that the lasing power can be ~60% efficiently coupled out along the photonic rib waveguides (Fig. 1d, e).

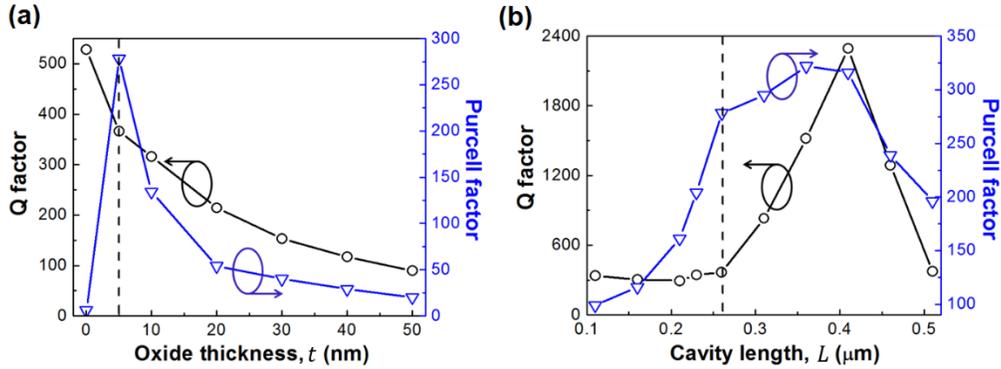

Fig. 2. Quality and Purcell factor dependency on (a) the oxide layer thickness, and (b) the laser cavity length, for the plasmonic combo-PCNB cavity. These maximum $Q$ and Purcell factors can be obtained by a dipole source placed at either $y = 100$ nm or $y = -100$ nm due to the symmetry in the cavity center. The dashed line indicates the optimized parameters based on a smallest cavity mode volume.

Our device requires 10 single-walled CNTs with pitch variation less than ~5 nm in the oxide layer. Experimentally we prefer to choose the separation method of CNT placement from solution due to the advantage of intending to select highly purified semiconducting single-walled CNTs and placing them onto a substrate with a specific pitch and orientation. Towards addressing the feasibility of placement of single chirality CNTs, here we can deploy the dielectrophoretic assembly method combining with polymer-mediated chirality sorting [39], showing an example of seven electrode pairs successfully bridged by an array of single chirality (9, 7) single-walled CNTs among the 10 electrode pairs. Note, the unbridged parts are caused by the nanotubes' length in the solution shorter than the electrode gap. The further experiment can be improved by narrowing the length distribution of CNTs in solution [39], such as by density gradient ultracentrifugation separating single-walled CNTs ranging in average length from <50 nm to ~2 μm [34].

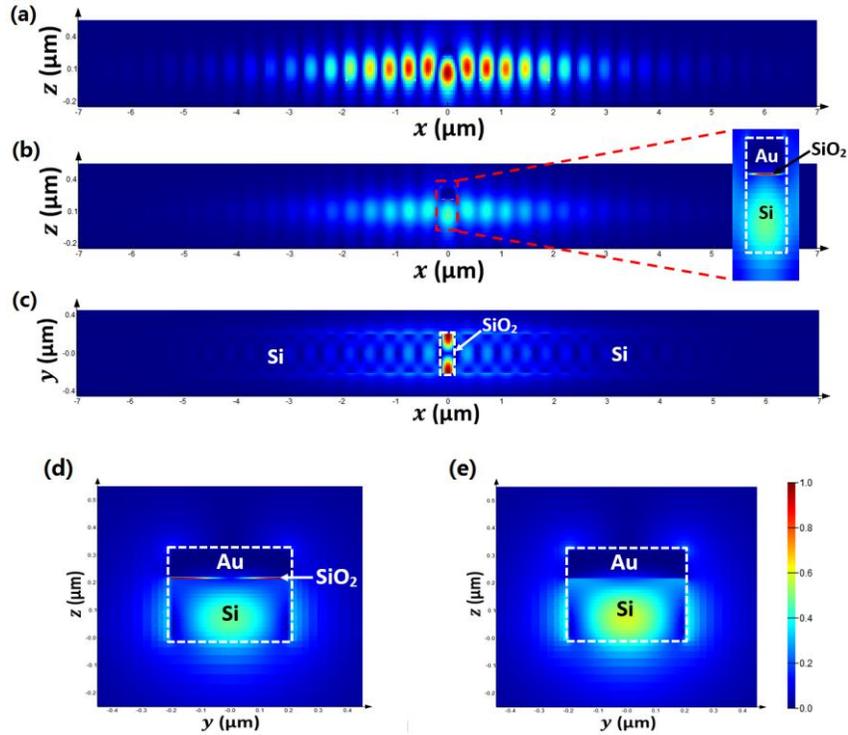

Fig. 3. Electric field profiles of a cavity mode (a) at the $xz$ plane along the $y = 0$ direction, (b) at the $xz$ plane along the $y = 100$ nm direction, Inset: field intensity of the active cavity region of the plasmonic hybrid mode, (c) at the $xy$ plane at $z = 217.5$ nm direction (i.e. within the thin oxide layer), (d) at the $yz$ plane along the $x = 0$ diection, for the MOS structure with $t=5$ nm oxide layer embedded in the PCNB cavity, respectively. (e) Electric field profiles at the $yz$ plane along the $x = 0$ diection for the MOS structure without the oxide layer embedded in the PCNB cavity. The cavity geometry with different materials is guided by a rectangular white dash lines. The color scale bar is normalized between 0 and 1.0.

The Purcell factor indicates the interaction strength between photons in the cavity and the laser gain medium by quantifying the spontaneous emission rate enhancement of an emitter inside a cavity. There are two methods to increase the Purcell factor [22], $F_p$, according to the widely used formula of $F_p = (6/\pi^2)(Q/V_n)$ [23], where $V_n$ is the diffraction-limited mode volume in a cubic half-wavelength in material, i.e. $V_n = V_{mode}/[(\lambda/2n_{cav})^3]$, $V_{mode}$ is the effective mode volume, $\lambda$ is the resonant free space wavelength of the cavity, and $n_{cav}(\lambda)$ is the effective cavity index. A rather classical approach is to enhance the cavity $Q$ factor [24]. However, this is somewhat unpractical due to the required increased wafer space and the lower modulation speed for lasers with high-$Q$ cavity (i.e. long photon lifetimes). The second possible approach is to decrease the $V_n$. Since $Q$ is ultimately limited in practice by these factors of bandwidth, material absorption, and fabrication tolerance, here we show that minimizing $V_n$ for a given $Q$ is a preferred solution. The internal dynamics leading towards the laser threshold are more efficiently utilized as the optical mode volume is smaller (i.e. higher $F_p$, and spontaneous emission coupling factor, $\beta$), and the smaller mode volume translates into a low pump power requirement to reach threshold [17, 19]. Here we find a relatively large $F_p$ for a reasonable $Q$ by scanning the oxide thickness and the cavity length, respectively (Fig. 2). Since $F_p$ depends on the polarization of dipole source excitation (i.e. z orientation is preferred) and the position of dipole source, we purposely place a dipole source at the peak of the electric field in the cavity (e.g. the position i or iii in Fig. 1c as we refer to Fig. 3c). A high Purcell factor of ~300, which is similar for a two-dimensional photonic crystal slab cavity [25], can be achieved due to the combo-cavity effect [19]. However the latter relies on a high $Q$ which introduces the aforementioned photon lifetime, footprint, and potential wavelength stabilization restrictions. Note, the maximum value of the LMI are observed at $t$=5 nm and $L$=260 nm, respectively, owing to the corresponding smallest cavity mode volumes (i.e. ~0.8 $(\lambda/2n_{cav})^3$) observed (dashed line Fig. 2a, b). Using this configuration, the plasmonic cavity exhibits a lasing peak wavelength of ~1522 nm (Fig. 1d). We conclude that a high Purcell factor can be achieved at a modest cavity $Q$, leading to a broader bandwidth and thus enabling broadband light sources with a high spontaneous emission rate [26], due to the relatively high coupling of CNTs emitter to the cavity.

## 3. Carbon nanotube laser performance

The Purcell effect enables the CNTs-based PCNB laser to significantly improve its performance via increasing the LMI, and hence the photon built-up efficiency (i.e. $\beta$-factor) inside the laser cavity. Here, we are particularly interested in the power output and the modulation speed characteristics of the Carbon-gain material driven laser. The steady state rate equations are utilized under continuous pumping without considering non-radiative recombination rate (Eq. 1, 2) [27], and the power output, $P_{out}$, is related to the photon number derived from the rate equations, yet associated with the other parameters from the previous optical simulation results (Eq. 3) [28].

$$\frac{dn}{dt} = \eta_i \frac{I}{qVol} - An - \beta\Gamma AS(n - n_0) \quad (1)$$

$$\frac{dS}{dt} = \beta An + \beta\Gamma AS(n - n_0) - \gamma S \quad (2)$$

$$P_{out} = \eta_c \frac{\alpha_m}{\alpha_m + \alpha_i} \frac{S_{ph}}{\tau_p} \frac{hc}{\lambda_o} V_{mode} \quad (3)$$

where $I$ is the injection current, and $\eta_i \frac{I}{qVol} = P$, $P$ is the pump rate, $\eta_i$ is the current injection efficiency, and we use the electroluminescence efficiency, $\eta_i$ =1.0×10$^{-4}$ of a CNT [6], $Vol$ is the active gain volume, here it is the volume of single-walled CNTs, $S$ is the photon number of a single lasing mode, $n$ is the carrier density, $A$ is the spontaneous emission rate and is enhanced by the Purcell effect via $A = F_p A_o$, where $A_o$ is the natural spontaneous emission rate of the material, and $A_o = 1/\tau_{sp0}$, $\tau_{sp0}$ is the spontaneous emission lifetime. Key for a fast gain re-modulation are the spontaneous emission lifetime, here of CNT, which is in the range of 20~200 ps [6], and the short photon lifetime of the plasmonic cavity ($\tau_p \propto Q$), and here $\tau_{sp0}$=100 ps. $\beta$ is the spontaneous emission coupling factor, $\Gamma$ quantifies the overlap between the spatial distribution of Carbon nanotube relative to a lasing mode, and $\Gamma = 5\%$ is estimated from the ratio between the area of 10 pieces of Carbon nanotube placed side by side and a ~200 nm$^2$ cross-section of a hybrid plasmonic mode. $\gamma$ is the total cavity loss rate per unit volume, $n_0$ is the carrier density at transparency, and $n_0 \approx 4.9\times10^{-13}$/cm$^3$ is used for chiral (9,2) Carbon nanotubes [29]. $\eta_c$ is the waveguide transmission efficiency of the PCNB cavity, $\alpha_m$ is the mirror loss, $\alpha_i$ is the intrinsic loss of the cavity, $\tau_p$ is the photon life time, and is proportional to the cavity $Q$ (i.e. $\tau_p = Q/(2\pi f)$, $f$ is the cavity resonant frequency), $h$ is the planck constant, $c$ is the light speed in vacuum, $\lambda_o$ is the lasing wavelength, $V_{mode}$ is the effective optical mode volume, and $S_{ph}$ is the photon density. Here we introduce a penetration length, $L_p$, into the PCNB cavity due to the undefined cavity length between the two Bragg mirror sections. The effective cavity length is $L_{eff} = L + 2L_p$, where $L_p$ can be written by [30],

$$L_p = -\lambda_o^2/(4\pi n_g)(\frac{\partial \phi}{\partial \lambda})_{\lambda_o} \tag{4}$$

$$Q = \frac{\pi}{1-R}[\frac{2Ln_g}{\lambda_o} - \frac{\lambda_o}{\pi}(\frac{\partial \phi}{\partial \lambda})_{\lambda_o}] \tag{5}$$

where $n_g$ is the group index of the cavity mode cycling in the resonator, the derivative of phase delay $(\partial \phi/\partial \lambda)_{\lambda_o}$ is related to the $Q$ factor (Eq. 5), $R$ is the modal reflectivity. Using both Eq. 4 and 5, $L_{eff}$ can be calculated and the effective cavity volume is thus evaluated by $Vol_{eff} = L_{eff}WH$. The photon density may be further estimated via $S_{ph} = S/Vol_{eff}$.

For the CNT laser we obtain the output power of about 3 (100) µW at a 1.0 (10) of the threshold pump rate (Fig. 4a). This is remarkable given the small gain volume, but can be understood by the high photon density (e.g. ~2×10$^{17}$/cm$^3$ at the threshold) inside the oxide layer of the laser cavity [19]. Below the threshold current (i.e. ~970 µA calculated for our case) the cavity laser behaves as an amplified spontaneous emission light source, showing the power output less than 3 µW with the injection current in the range of 0~1000 µA (inset Fig. 4a).

Higher modulation frequencies of directly-modulated semiconductor lasers allowing larger data rates are desired in relatively short-distance data transmissions. However, conventional semiconductor laser sources have their bandwidths limited to around 40 GHz due to gain compression effects and parasitic electrical capacitances. The 3-dB role off modulation bandwidth ($f_{3dB}$, defined as the frequency at which the response function decays to half of its zero-frequency value) is estimated through the small signal response (direct modulation) of the CNT laser by observing the spectral response function [31],

$$\frac{H(\omega)}{H(0)} = \frac{\omega_r^2}{\sqrt{(\omega^2-\omega_r^2)^2+\omega^2\omega_p^2}} \tag{6}$$

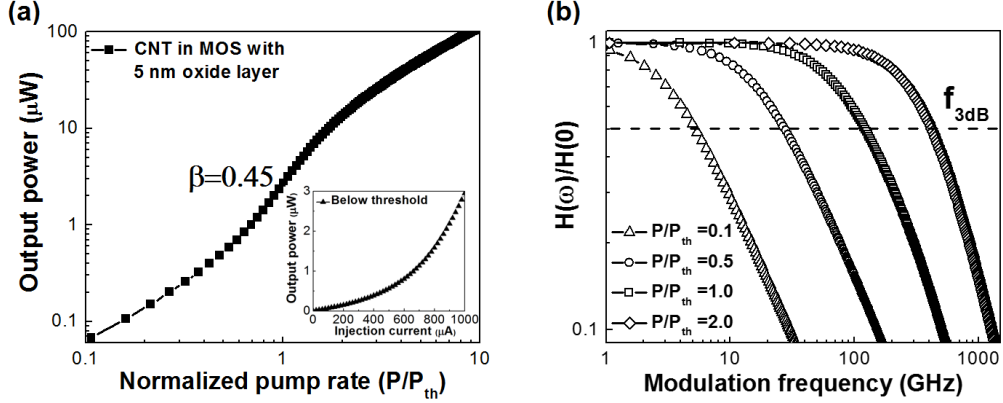

Fig. 4. (a) Laser output power as a function of normalized pump rate (i.e. $P/P_{th}$) for the PCNB cavity based CNT light source, exhibiting a high β factor ~0.45 which helps to reduce the laser threshold. Here 5 nm thickness oxide layer is used in the combo cavity. The inset shows the output power dependence on the injection current below threshold. (b) Modulation bandwidth of the single-walled CNTs laser/light emitting diode with below, equal, and above threshold pump rate, $P_{th}$. A ~150 GHz modulation frequency is calculated at a 3 dB bandwidth with the pump rate of $P/P_{th} = 1.0$.

where $\omega$ is the optical cavity angular frequency, $\omega_p = \gamma_c + \Gamma_T(1 - \beta N_0 + \beta S_0)$, and $\omega_r^2 = \Gamma_T[\gamma_c(1 + \beta S_0) - \beta(1 - \beta)\Gamma_T N_0]$, $S_0$ and $N_0$ are the steady-state photon number and population inversion number, respectively, and $\gamma_c = 1/\tau_p$, $\Gamma_T$ is the transition rate of excited state population, which is equal to the spontaneous emission rate, $A$. The frequency response of the device below lasing threshold is also calculated using spontaneous emission lifetime, which is equal to carrier lifetime but neglecting non-radiative recombination lifetime. Here we theoretically show the frequency response of the device with up to 2 times of the threshold pump rate, delivering a 3-dB bandwidth of more than 100 (400) GHz at a 1.0 (2.0) of the threshold pump rate (Fig. 4b). The modulation bandwidth increases with higher injection current, which can be understood as an interplay between photonic and electronic rates of both the cavity and the external pump (i.e. driving current). If the internal laser cavity is fast enough, the higher pump rate drives the gain medium faster into population inversion. Given the lossy plasmonics cavity, this inversion is rapidly depleted and hence can be re-excited more promptly compared to larger diffraction limited devices.

## 4. Sensitivity to fabrication imperfections

The deviation in hole size can lead to various effects on the optical properties of the photonic crystal laser. Firstly, it can cause shifts in the position of the photonic bandgap, altering the range of wavelengths that the laser can efficiently emit or reflect. This shift can have implications for the laser's output characteristics and operating parameters. Furthermore, the broadening of the bandgap due to hole size deviations can limit the laser's ability to effectively block specific wavelengths, affecting its performance in applications requiring precise control over the emission spectrum. Additionally, altered dispersion characteristics and group velocity can impact the laser's propagation properties, leading to dispersion-induced broadening and changes in pulse characteristics.

To mitigate the effects of hole size deviations, it is crucial to implement tighter fabrication tolerances during the manufacturing process. This requires advancements in fabrication techniques, such as lithography and etching, to achieve greater precision in hole dimensions.

Design optimization that considers anticipated hole size deviations can also help in minimizing their impact. Post-fabrication characterization plays a vital role in assessing the extent of hole size deviations and their influence on the laser's performance. Understanding and addressing hole size deviations are critical steps in achieving the desired optical properties and optimal performance in 1D photonic crystal lasers. By refining fabrication techniques, optimizing design parameters, and conducting thorough characterization, we can enhance the reliability and functionality of photonic crystal lasers for various applications.

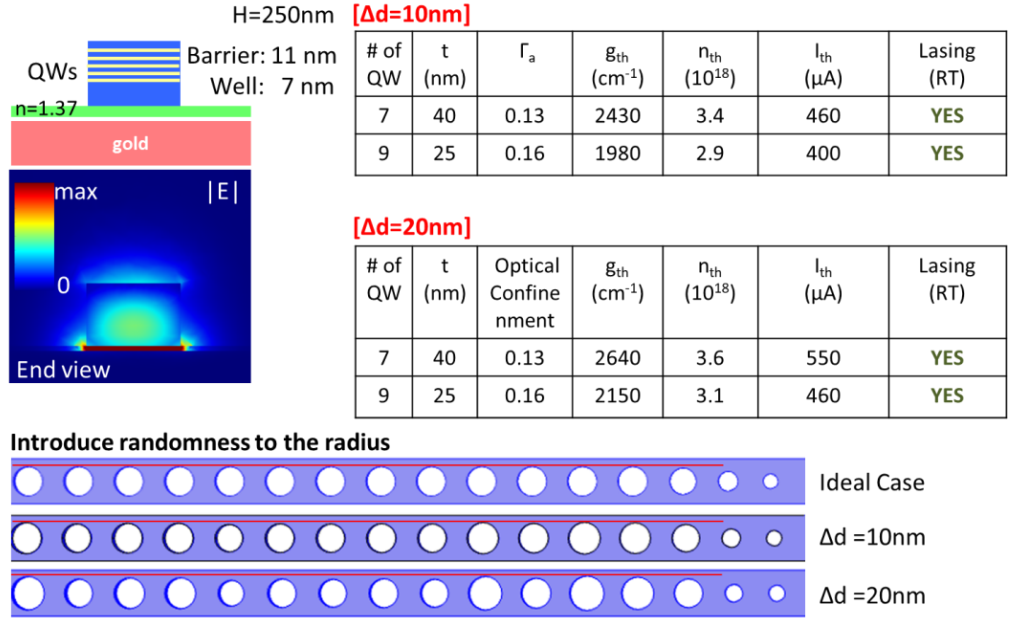

Fig. 5. Illustrates the effect of random imperfections in nanoholes on the performance of an electrically driven nanobeam laser. Despite a significant variation of 20 nm in hole size from the ideal design, lasing at room temperature is expected, highlighting the robustness of the laser against these imperfections.

## 5. Conclusion

In conclusion, our theoretical investigation focuses on plasmonic photonic crystal hybrid lasers utilizing Carbon nanotubes (CNTs) as the gain material. These hybrid lasers offer significant potential as high-performance on-chip light sources for telecom applications. Simulation results demonstrate that the hybrid HPP (Hybrid Plasmon-Photon) waveguide mode, generated from the emission of CNTs, exhibits strong coupling efficiency of approximately 60% with the 1-D photonic crystal cavity lasing mode. This coupling enables efficient energy transfer and enhanced light emission within the device. Compared to gain compression-limited devices, our proposed light source enables faster modulation due to two key factors: the strong Purcell effect, resulting in an enhancement factor of approximately 300 ($F_p$), and the short spontaneous emission lifetime of CNTs. These combined characteristics allow for modulation speeds of hundreds of GHz with a 3dB roll-off, along with tens of microwatts of optical power above the laser threshold. The integration of CNT internal processes with the plasmonic cavity architecture presents an alternative approach for the realization of active components in next-generation photonic circuits. The monolithic integration schemes proposed in this study offer a promising path for the development of advanced photonic devices, enabling improved performance, faster modulation speeds, and higher optical power output.